\documentclass[twocolumn,amsmath,amssymb]{revtex4-1}

\usepackage{graphicx}% Include figure files
\usepackage{dcolumn}% Align table columns on decimal point
\usepackage{bm}% bold math
\usepackage{indentfirst}

\usepackage{float}
\usepackage{amsfonts}
\usepackage{caption}
\usepackage{subfigure}
\usepackage{multirow}
\usepackage{amsmath}
\usepackage{amsfonts}
\usepackage{amssymb}
\usepackage{mathrsfs}
\usepackage{color}
\usepackage{amsthm}

\def\I{\mathbf{I}}
\def\Q{\mathbf{Q}}
\def\P{\mathbf{P}}
\def\I{\mathbf{I}}

\def\n{\mathbf{n}}
\def\m{\mathbf{m}}

\def\x{\mathbf{x}}

\def\d{\mathrm{d}}

\def\_v{\mathbf{v}}

\def\S{\mathbf{S}}

\begin{document}

\preprint{APS/123-QED}
\title{Neural Network-Based Tensor Model for Nematic Liquid Crystals with Accurate Microscopic Information}
%\title{Tensor Models for Liquid Crystals Incorporating Molecular-Level Information Using Neural Networks}
%\title{Novel Neural Network Based Micro-Macro models for Liquid Crystals}
%\title{Molecular-level Tensor Models for Nematic Liquid Crystals Using Neural Networks}
\author{Baoming Shi$^1$}
\author{Apala Majumdar$^{2}$}
\author{Lei Zhang$^{3}$}
\email{Contact author: zhangl@math.pku.edu.cn}

\affiliation{
   $^{1}$School of Mathematical Sciences, Peking University, Beijing 100871, China.\\
   $^2$Department of Mathematics and Statistics, University of Strathclyde, G1 1XQ, UK.\\
   $^3$Beijing International Center for Mathematical Research, Center for Quantitative Biology, Center for Machine Learning Research, Peking University, Beijing 100871, China.
}%

\begin{abstract}
The phenomenological Landau-de Gennes (LdG) model is a powerful continuum theory to describe the macroscopic state of nematic liquid crystals. However, it is invariably less accurate and less physically informed than the molecular-level models due to the lack of physical meaning of the parameters.
We propose a neural network-based tensor (NN-Tensor) model for nematic liquid crystals, supervised by the molecular model. Consequently, the NN-Tensor model not only attains energy precision comparable to the molecular model but also accurately captures the Isotropic-Nematic phase transition, which the LdG model cannot achieve. The NN-Tensor model is further embedded in another neural network to predict liquid crystal configurations in a domain-free and mesh-free manner. We apply the NN-Tensor model to nematic liquid crystals in a number of two-dimensional and three-dimensional domains to demonstrate it can efficiently identify rich liquid crystal configurations in both regular and non-regular confinements.

\end{abstract}

%\keywords{liquid crystal, nematic liquid crystal, Landau-de Gennes model, Maier-Saupe energy, neural network}
%Use showkeys class option if keyword
                              %display desired
\maketitle

Nematic liquid crystals (NLCs) are classical mesophases that combine fluidity with long-range orientational order, but have no positional order \cite{de1993physics}. The NLC molecules tend to align along certain locally preferred directions referred to as nematic ``directors'' in the literature. Consequently, NLCs have anisotropic or direction-dependent physical, optical and rheological properties \cite{sonin2018pierre,stewart2019static}, making them the working material of choice for electro-optic devices, photonics, actuators, elastomers and with tremendous potential in biophysics, nanotechnologies and healthcare technologies  \cite{LAGERWALL20121387,bisoyi2021liquid,loussert2013manipulating}. 

There are several mathematical theories for NLCs, ranging from detailed atomistic models, molecular models to fully macroscopic models \cite{de1993physics,wang2021modeling,doi1988theory}. The molecular models are characterized by a probability distribution function for the molecular orientations, known as the orientation distribution function (ODF) and intermolecular interaction kernels that contain information about molecular shapes, sizes and interactions  \cite{yin2022solution,zhang2012onsager,liang2014rigid,liu2005axial}. However, molecular models suffer from the curse of high dimensionality. On the other hand, macroscopic models are phenomenological and computationally efficient, with fewer degrees of freedom. In particular, the celebrated Landau-de Gennes (LdG) theory has been widely used for modelling two-dimensional (2D) and three-dimensional (3D) liquid crystal equilibria \cite{wu2023topological,yin2020construction,hu2016disclination,muvsevivc2006two,shi2022nematic}. The LdG model describes the macroscopic NLC state by a $\mathbf{Q}$-tensor order parameter, which contains information about the nematic directors and the degree of orientational ordering within its eigenvectors and eigenvalues respectively. However, the LdG free energy is phenomenological, that is, the precise physical meaning of the LdG parameters are not relatable to molecular-level quantities. The LdG theory is strictly speaking, only valid near the Isotropic-Nematic (I-N) phase transition temperature, and does not satisfy physical constraints at low temperatures \cite{ball2010nematic}.

A more physically informed macroscopic model would be derived from the molecular models \cite{han2015microscopic,mei2015molecular,ball2010nematic}. One can approximate the ODF by its second moment, $\Q$, and the corresponding energy can be written in terms of $\Q$. This procedure has been applied to nematic phase and other LC phases with complex molecular shapes  \cite{xu2018tensor}. However, this procedure often results in complex functionals with implicit-dependence on $\Q$, making them computationally unwieldy and perhaps of interest only in one-dimensional settings \cite{han2015microscopic,mei2015molecular}. The quasi-entropy method in  \cite{xu2022quasi} combines molecular-level information with an explicit $\Q$-dependence, although it does not achieve molecular-level precisions. The trade-off between computational efficiency and accuracy has been a major dilemma in computational liquid crystal physics. %he phenomenological tensor model (e.g. LdG model and quasi-entropy based tensor model) is computationally cheap but not so accuracy, and the physical tensor models fully derived from molecular model are computationally expensive. 

Recent deep learning advances are capable of bridging the gap between accuracy and computational efficiency. For instance, Behler-Parrinello neural network \cite{behler2007generalized} and deep potential \cite{zhang2018deep,zhang2021phase} use deep neural networks to build molecular dynamics supervised by first-principle in order to achieve the accuracy of quantum mechanics. 

In this letter, we propose a computationally flexible neural network-based tensor (NN-Tensor) model, which employs the microscopic molecular model to guide a new trained bulk energy density for NLCs. Consequently, the NN-Tensor model not only accurately captures the Isotropic-Nematic phase transition but also preserves energy precision comparable to the molecular model, which the LdG model cannot achieve. The NN-Tensor model is further embedded in another neural network to efficiently compute liquid crystal configurations in a domain-free and mesh-free manner.

%The total free energy is a LdG-type energy comprising the trained bulk energy density and a conventional one-constant elastic energy density to penalize spatial inhomogeneities. 

%The NN-Tensor model is embedded in another neural network to determine the stable nematic liquid crystal configurations in prototype geometries. We demonstrate the broad applicability of the NN-Tensor model by applying it on both spatially homogeneous and inhomogeneous problems.

%In fact, neural networks have been used in deriving micro-to-macro models, e.g. Behler-Parrinello neural network \cite{behler2007generalized}, deep tensor neural network \cite{schutt2017quantum}, and deep potential \cite{zhang2018deep,zhang2021phase}, which use neural networks to represent the potential energy supervised by first-principle. For NLCs, we employ the Onsager molecular model to guide a new trained bulk energy density in NN-Tensor model. Thus, NN-Tensor model 

%NEED TO REVISE
%This letter spearheads the use of NN-Tensor model for the analysis of 2D and 3D NLC equilibria, for both spatially homogeneous and inhomogeneous problems. employs the Onsager molecular model with the Maier-Saupe intermolecular interaction potential to guide a new trained bulk energy density for nematic liquid crystals. The total free energy is a LdG-type energy comprising the trained bulk energy density and a conventional one-constant elastic energy density to penalize spatial inhomogeneities. 

%Results
In the molecular model, the order parameter is the ODF, $f(\m)$, which measures the probability of finding molecules with orientation $\m \in \S^2$ \cite{onsager1949effects}. For example, the Maier-Saupe (M-S) energy \cite{maier1958einfache} is given by:
\begin{equation}
\begin{aligned}
&E_M(f)=\tau \int_{\S^2} f(\m)\ln f(\m)\d \m \\
&+ \frac{\kappa}{2} \int_{\S^2} \int_{\S^2}\left(\frac{1}{3}-(\m \cdot\hat{\m})^2\right)f(\m)f(\hat{\m})\d\m \d \hat{\m}.
\label{onsager model}
\end{aligned}
\end{equation}
The first term, $\tau E_{entropy}(f)$, is the entropy energy with a positive $\tau$. $\tau$  is proportional to the absolute temperature. The second term, $\kappa E_{potential}(f)$, is the potential energy induced by M-S two-body interactions and $\kappa>0$ measures the strength of the intermolecular interactions. We set $\tau=1$ in \eqref{onsager model}
 and adjust the functional with the parameter $\kappa$, so that $\kappa$ is inversely proportional to the absolute temperature. The computational domain is $\Omega \times \S^2$ for inhomogeneous problems, which is two dimensions higher than $\Omega$ and hence, the curse of high dimensionality. 
\begin{figure*}
    \centering
    \includegraphics[width=0.9\linewidth]{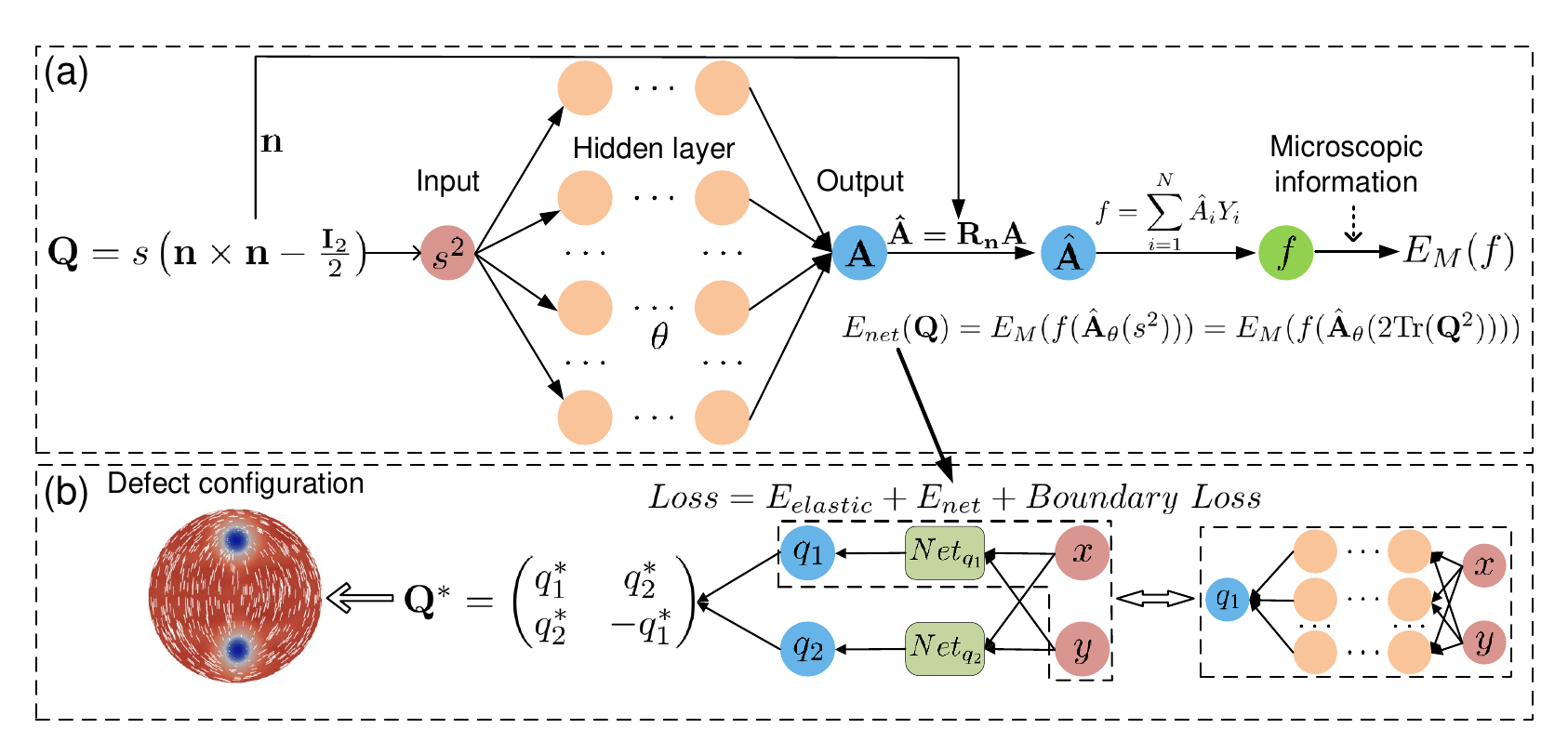}
    \captionsetup{justification=raggedright}
    \caption{(a) The 2D NN-Tensor model with microscopic information. The parameters that need to be trained are represented by $\theta$. The 3D NN-Tensor model can be seen in Supplementary Material. (b) The NN-Tensor model is embedded in the loss function of the second neural network to find the stable configurations.  
    %The spatial coordinates of the confined domain are \((x, y)\).
    }
    \label{figure: schematic 2D}
\end{figure*}

The second moment of $f$,
\begin{equation}
\Q(x)=\int_{\mathbb{S}^2} \left(\m \times \m-\frac{\I_3}{3}\right) f(\x,\m) \d \m,
\label{Q-tensor}
\end{equation}
is often used as a measure of the orientational ordering. It is a traceless $3\times 3$ matrix by definition, so that $\Q=0$ for the uniform ODF without any special or distinguished directions. From the spectral decomposition theorem, we have that
$\Q = \sum_{i=1}^{3} \lambda_{i} \n_i \times \n_i$,
where $\left\{ \n_1, \n_2, \n_3 \right\}$ are the eigenvectors  and $\lambda_1\leqslant\lambda_2\leqslant\lambda_3$ are the associated eigenvalues respectively, subject to $\sum_{i=1}^{3}\lambda_i = 0$. Further, the eigenvalues are naturally constrained to be
\begin{equation}
-\frac{1}{3} < \lambda_i=\int_{\S^2}\left(\m \cdot \n_i \right)^2 f(\x,\m)\d \m -\frac{1}{3}<\frac{2}{3}.
\label{domain of eigenvalue}
\end{equation} by the second moment definition of $\Q$ in \eqref{Q-tensor}. A $\Q$ is deemed admissible in the molecular framework if and only if $\lambda_i \in \left(-\frac{1}{3}, \frac{2}{3} \right)$ for $i=1,2,3$. % the inequalities \eqref{domain of eigenvalue} are satisfied. % in the moleculaif $\Q$ is of the formNot all  symmetric and traceless $3\times 3$ $\Q$ tensor can be derived from \eqref{Q-tensor}. Actually, if a $\Q$ tensor has the form in  \eqref{Q-tensor}, multiplying both sides in \eqref{Q-tensor} by $\n_i$, the eigenvalues are naturally constrained to be

We work in a variational framework and a viable free energy, $E(\Q)$, must contain at least two competing contributions: a bulk energy, $E_b(\Q)$, which dictates the preferred NLC phase in spatially homogeneous samples, and the
elastic energy, $E_{elastic}(\Q,\nabla \Q)$, which penalizes spatial inhomogeneities \cite{de1993physics}. The free energy must also respect frame indifference and material symmetry properties i.e., for any $\P \in SO(3)$,
\begin{equation}
E_b(\P \Q \P^T)=E_b(\Q).
\label{frame indifference}
\end{equation}
The famous LdG energy, $E_{LdG}(\Q_M)$ is a simple and typical free energy that meets these requirements \cite{de1993physics}, where $\Q_M$ is the LdG tensor order parameter. However, $\Q_M$ is independent of the ODF and is not compatible with the inequalities \eqref{domain of eigenvalue}. Moreover, the LdG bulk energy, $E_b(\Q_M)$, is a quartic polynomial or a truncated expansion near the isotropic state, $\Q_M=0$, whose coefficients do not have a clear molecular interpretation. Hence, it is not surprising that LdG predictions can fail for low temperatures or near defects and critical phenomena. %in \cite{ball2010nematic}, the following bulk energy is derived from the molecular model with M-S potential energy
%\begin{equation}
%E_b(\Q)=
%\tau\left(\underset{f\in \mathcal{A}_{\Q}}{\mathrm{inf}}\int_{\S^2}f(\m)\ln{f(\m)\d\m}-\kappa|\Q|^2\right),
%\label{bulk from M-S}
%\end{equation}
%where $\mathcal{A}_{\Q}$ is the set of all distributions $f$ whose second moment \eqref{Q-tensor} is $\Q$. With this bulk energy, the corresponding energy minimizers satisfy the physical constraints \eqref{domain of eigenvalue}, since $E_b(\Q)\rightarrow \infty$ if $\lambda_1(\Q) \rightarrow -1/3$ \cite{ball2010nematic}. However, the bulk energy in \eqref{bulk from M-S} is implicitly dependent with $\Q$.
% The bulk energy in the LdG energy is a quartic polynomial in the tensor. However, there are no physical constraints on the LdG tensor compatible with the constraints \eqref{domain of eigenvalue}.

%To obtain a more reliable tensor model, a natural method is to derive a tensor based free energy from a molecular energy \cite{han2015microscopic,mei2015molecular,ball2010nematic}. For example, 

\emph{NN-Tensor model}--%Considering the ability of neural networks to learn complex potential energy,
Our key idea is to use neural networks to map a given $\Q$ to an ODF, and then to the M-S energy \eqref{onsager model}. For brevity, we first consider the 2D case in Figure \ref{figure: schematic 2D}(a). The 2D tensor order parameter is of the form $\Q_{2D}=s \left(\n \times \n -\I_2/2\right)$, where $\n \in \S^1$ and $0 \leqslant s< 1$ \cite{han2019transition}. We use the Bingham closure method \cite{chaubal1995comparison,chaubal1998closure} to define an ODF, $f_{\Q_{2D}}(\m)$, for a given $\Q_{2D}$ with $0\leqslant s<1$, where%  The key point for deriving tensor model from the molecular model is to approximate the orientation distribution function $f(\m)$ by a suitable
%function of $\Q$, referred to as closure method \cite{chaubal1995comparison,chaubal1998closure}. Here we use the Bingham closure
\begin{equation}
f_{\Q_{2D}}(\m)=\frac{\text{exp}(\mathbf{B}_{\Q_{2D}}:\m\times \m)}{\int_{\S^1} \text{exp}(\mathbf{B}_{\Q_{2D}}:\m\times \m)}, \quad \m \in \S^1;
\label{2D bingham distribution}
\end{equation}
$\mathbf{B}_{\Q_{2D}}$ is a symmetric traceless $2\times2$ matrix implicitly determined by $\Q_{2D}$ \cite{li2015local}. Next, we can generate the data and the corresponding label $(\Q_{2D},f_{\Q_{2D}})$, but it is necessary to convert the label $f_{\Q_{2D}}$
into a finite-dimensional $N\times 1$ vector. One approach is to discretize it into $f_{\Q_{2D}}(\m_j)$, where $\{\m_j\},1\leqslant j \leqslant N$, are the discretized points on the unit circle $S^1$. However, this would lead to an excessively large number of neurons in the output layer. A more efficient method is to use the discrete spherical harmonics transform \cite{mohlenkamp1999fast},  $\mathbf{A}$, which can well approximate 
$f_{\Q_{2D}}$ with a relatively small $N$ \cite{shen2011spectral}. Further, the spherical harmonics transform is rotationally invariant, that is, if we rotate $f_{\Q_{2D}}(\m)$ to $f_{\Q_{2D}}(\P_{\n} \m), \P_{\n} \in SO(2)$, then the rotated expansion coefficients can be calculated from $\hat{\mathbf{A}}=\mathbf{R}_{\n} \mathbf{A}$ (see Supplementary Material). Thus, we only sample $\Q_{2D}^i=\text{diag}(s_i/2,-s_i/2)$ for $ 0 \leqslant s_i<1$, with a fixed eigenvector $\n_{e}=(1,0)$, and the rest follows from rotational invariance. %expansion coefficients of $f_{\Q}$ for which $\n(\Q)\neq (1,0)$ can be directly calculated. 
To maintain frame indifference as in \eqref{frame indifference}, we encode $\Q_{2D}$ as a function of its non-negative eigenvalue $s/2$. We specifically encode $\Q_{2D}$ as $s^2$ because $s^2=2\textbf{Tr}(\Q_{2D}^2)$ simplifies the differentiation of $\Q_{2D}$. After training the NN-Tensor model, we obtain the following bulk energy: 
\begin{equation}
E_{net}(\Q_{2D})=E_M(f(\hat{\mathbf{A}}_\theta(s^2))), s^2=2\text{Tr}(\Q_{2D}^2),
\label{eq: trained bulk energy}
\end{equation}
where $\theta$ denotes the set of parameters in the neural network and  $E_M$ is the molecular free energy in \eqref{onsager model}.

For 3D NN-Tensor model, the number of 3D discrete spherical harmonics coefficients is much larger than in the 2D case, which results in an excessive number of neurons in the output layer. Therefore, we  replace fitting the map between $\Q$ and $f_\Q$ with directly labeling $\Q$ by $E_{entropy}(f_\Q)$ and $E_{potential}(f_\Q)$ respectively (see Supplementary Material).  We ensure frame indifference by encoding $\Q$ as a function of its minimum eigenvalue, $\lambda_1$, and second minimum eigenvalue, $\lambda_2$. Specifically, we encode $\Q$ as $\sum_{i=1}^3 \lambda_i^2=\textbf{Tr}(\Q^2)$ and $\sum_{i=1}^3 \lambda_i^3=\textbf{Tr}(\Q^3)$ for convenient differentiation, and they are bijective with respect to $\lambda_1$ and $\lambda_2$ within the domain of definition specified in \eqref{domain of eigenvalue}. The 3D NN-Tensor model can be trained to yield the following bulk energy as a function of 3D $\Q$ tensor defined in \eqref{Q-tensor}: 
\begin{equation}
E_{net}(\Q)=E_M\left(Net_\theta\left(\sum_{i=1}^3 \lambda_i^2,\sum_{i=1}^3 \lambda_i^3\right)\right).
\label{eq: 3D trained bulk energy}
\end{equation}

\begin{figure*}
    \centering
    \includegraphics[width=0.90\linewidth]{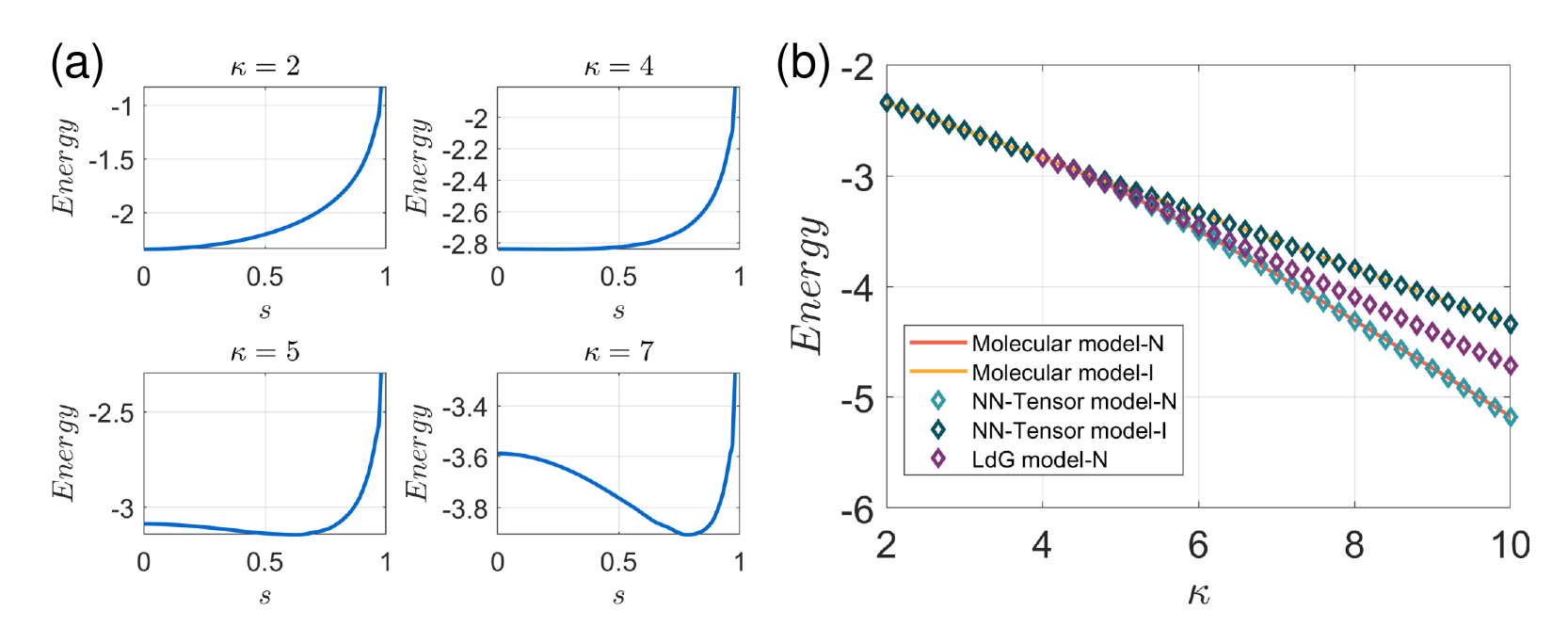}
    \captionsetup{justification=raggedright}
    \caption{(a)$E_{net}(\Q_{2D})=E_M(f(\hat{\mathbf{A}}_\theta(s^2))$ versus $s$ at $\kappa=2,4,5,7$. (b) The energies of the isotropic (I) and nematic (N) phases versus $\kappa$. The energy of isotropic phase in the NN-Tensor model and molecular model are $E_{net}(\Q=0)$ and $E_M(f\equiv 1/2\pi)$, respectively. For $\kappa>4$, the energy of the nematic phase in the NN-Tensor model is $\min_{0\leqslant s <1}E_M(f(\hat{\mathbf{A}}_\theta(s^2))$, and in the molecular model, it can be analytically calculated   \cite{liu2005axial}.}
    \label{phase transition 2D}
\end{figure*}

\emph{Phase transitions}--We first investigate whether the trained bulk energies in \eqref{eq: trained bulk energy} and \eqref{eq: 3D trained bulk energy} can accurately capture the I-N phase transition. In the 2D case, we plot the energy $E_{net}(\Q_{2D})$ in \eqref{eq: trained bulk energy}  versus $s$ for various values of $\kappa$ in Figure \ref{phase transition 2D}(a). For $\kappa=2$, the temperature is high and and the entropy energy in \eqref{onsager model} dominates, so that the isotropic phase $s=0$ ($\Q=0$) is the global energy minimizer. For $\kappa=4$, the isotropic phase nearly loses its stability, which is consistent with the fact that molecular free energy in \eqref{onsager model} undergoes a second-order I-N phase transition at $\kappa^*=4$ \cite{liu2005axial}. For $\kappa=5$, the isotropic phase loses stability and the nematic phase with $s\approx 0.64$ achieves the energy minimum. As $\kappa$ further increases to 7, the nematic phase with $s\approx 0.79$ achieves the energy minimum, indicating that the nematic phase becomes increasingly orientationally ordered as the temperature further decreases. In Figure \ref{phase transition 2D}(b), we plot the energies of the isotropic and nematic phases with $E_{net}(\Q_{2D})$ and the M-S free energy \eqref{onsager model}, demonstrating excellent agreement between the two approaches. In contrast, the LdG model is only accurate near $\kappa^*$ and fails to provide an accurate approximation at low temperatures.

\begin{figure}
        \centering
\captionsetup{justification=raggedright}    \includegraphics[width=0.96\linewidth]{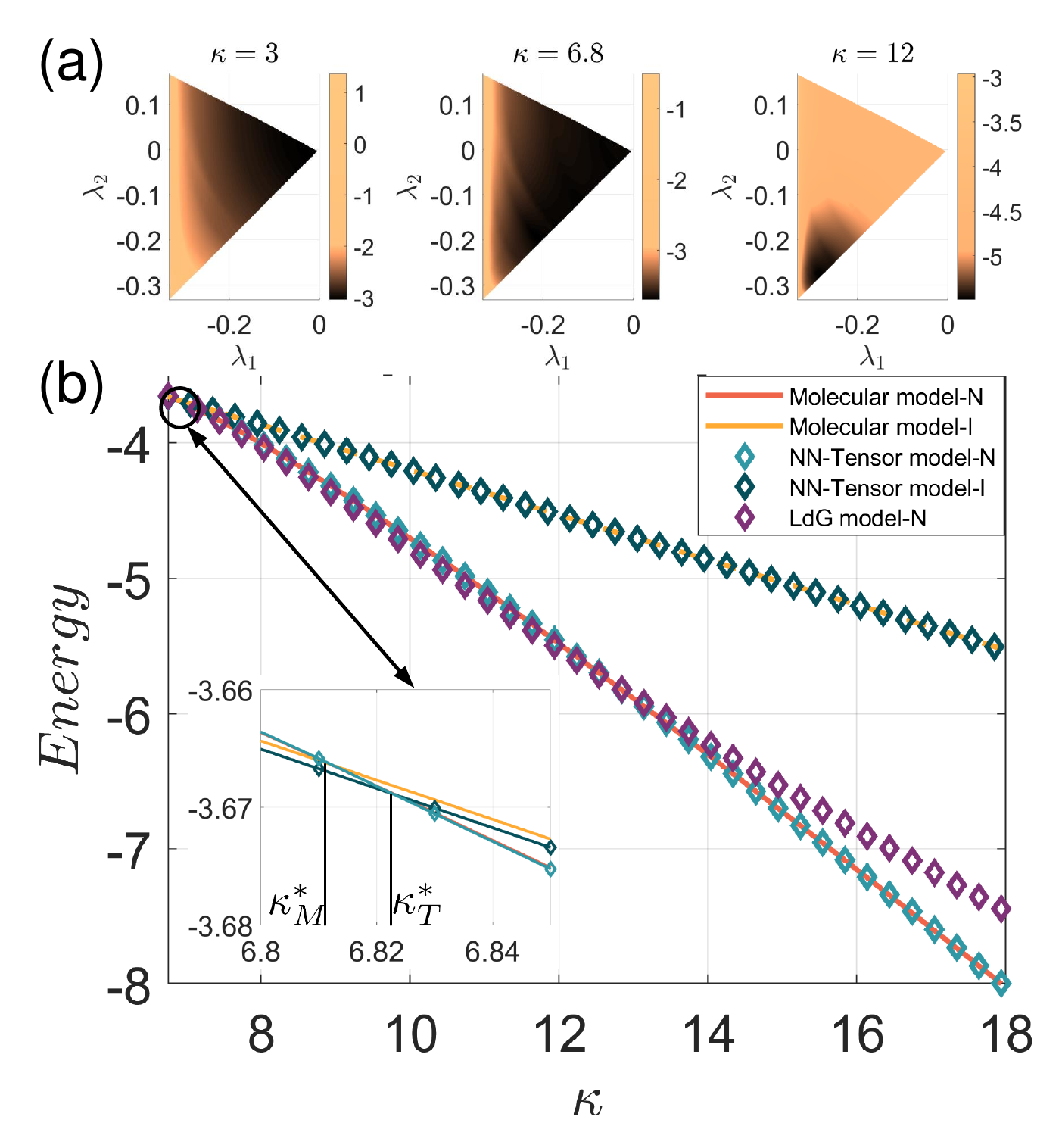}
    \caption{(a) $E_{net}(\Q)$ versus  $\lambda_1$, $\lambda_2$ at $\kappa=3$, $6.8$, $12$. (b) The energies of isotropic and nematic phases in the molecular model and NN-Tensor model versus $\kappa$. $K^*_M$ and $K^*_T$ are the I-N phase transition temperatures in the molecular model and NN-Tensor model, respectively.}
    \label{phase transition 3D}
\end{figure}

In the 3D case, we plot the energy in \eqref{eq: 3D trained bulk energy} versus $\lambda_1,\lambda_2$, for different values of $\kappa$ in Figure \ref{phase transition 3D}(a). For $\kappa=3$, the isotropic phase $\Q=0$ ($\lambda_1$=$\lambda_2$=0) is the global minimizer. At $\kappa=12$, the isotropic phase loses  stability and the nematic phase with $\lambda_1=\lambda_2\approx-0.28$ achieves the energy minimum. At $\kappa=6.8$, there are two energy minimizers: the isotropic phase and the nematic phase (with $\lambda_1=\lambda_2\approx -0.132$). Hence, the 3D NN-Tensor model predicts a first-order I-N phase transition, consistent with the molecular model which predicts a first-order I-N phase transition for $\kappa^* \in (6.7,7.5)$ \cite{liu2005axial,yin2022solution}. In Figure \ref{phase transition 3D}(b), the NN-Tensor model fits the molecular model well both near the phase transition temperature, $\kappa_M^*$, and for lower temperatures. 

\emph{Confined NLC systems}--For confined spatially inhomogeneous systems, we add a one-constant elastic energy density \cite{ball2010nematic} to the trained bulk energy density to define a total free energy, as shown below:
\begin{equation}
E(\Q)=\int_\Omega \frac{1}{2}|\nabla \Q|^2 + \lambda^2 E_{net}(\Q) \d \mathbf{x},
\label{eq: Energy 2D inhomogeneous}
\end{equation}
where $\Omega$ is the rescaled 2D or 3D domain and $\lambda$ is a scaling parameter that contains information about the geometrical size and macroscopic material properties. In particular, $\lambda$ is proportional to domain size and large values of $\lambda$ are experimentally relevant \cite{canevari2017order,shi2023hierarchies}. The physically observable configurations are modelled by energy minimizers 
subject to the imposed boundary conditions. The end-to-end representation of the bulk energy density, $E_{net}(\Q)$ is convenient for taking the derivative with respect to $\Q$, which in turn, allows us to find energy minimizers using the gradient descent method. For computational efficiency and flexibility, we use two (five) neural networks, $Net_{q_i}$, to approximate two (five) elements of the 2D (3D) $\Q$-tensor (see Figure \ref{figure: schematic 2D}(b)). Numerical tests in the Supplementary Material demonstrate that it reaches minima of \eqref{eq: Energy 2D inhomogeneous} faster than classical numerical methods, such as finite difference methods. This neural network approach is domain-free and mesh-free, making it particularly suitable for complex and experimentally relevant domains \cite{monderkamp2021topology,liang2011nematic,noh2020PRR}. The minimizer of \eqref{eq: Energy 2D inhomogeneous}, $\Q_{\theta^*}(x,y)$, is obtained by training $Net_{q_i}$ with the following loss function,

%For comparison purposes, the LdG bulk energy is a quartic polynomial with analytic derivatives and the NN-Tensor model approach, which employs $E_{net}(\Q)$, is relatively computationally expensive. Therefore, we use two (five) neural networks, $Net_{q_i}$, to approximate two (five) elements of the 2D (3D) $\Q$-tensor (see Figure \ref{figure: schematic 2D}(b)). The minimizer of \eqref{eq: Energy 2D inhomogeneous}, $\Q_{\theta^*}(x,y)$, is obtained by training $Net_{q_i}$ with the following loss function,
\begin{equation}
\begin{aligned}
&Loss(\theta)=\underbrace{\frac{W}{N_b}\sum_{i=1}^{N_{b}}\left|\Q_\theta(x^b_i,y^b_i)-\Q_{bc}(x^b_i,y^b_i)\right|^2}_{Boundary \text{ }Loss}+\\
&\underbrace{\frac{1}{2N_{in}}\sum_{i=1}^{N_{in}} |\nabla \Q_\theta(x^{in}_i,y^{in}_i)|^2}_{E_{elastic}}+\underbrace{\frac{\lambda^2}{N_{in}}\sum_{i=1}^{N_{in}} E_{net}(\Q_\theta(x^{in}_i,y^{in}_i))}_{E_b=\lambda^2E_{net}},
\end{aligned}
\end{equation}
where $(x_i^{in},y_i^{in})\in \Omega \backslash \partial \Omega, i=1,\cdots,N_{in}$ are uniformly sampled points in $\Omega$, and $(x_i^{b},y_i^{b})\in \partial \Omega, i=1,\cdots,N_{b}$ are uniformly sampled points on $\partial \Omega$. The boundary loss with penalty weight, $W$, enforces preferred boundary conditions, represented by $\Q_{bc}$. Examples of preferred boundary conditions include 2D tangential boundary conditions, $\Q_{bc}(x_i^b,y_i^b)=-s_+(\vec{\nu}_i^b \times \vec{\nu}_i^b -\I_2/2)$, and 2D homeotropic boundary conditions, $\Q_{bc}(x_i^b,y_i^b)=s_+(\vec{\nu}_i^b \times \vec{\nu}_i^b -\I_2/2)$, where $\vec{\nu}_i^b$ is the outer normal vector of $\partial \Omega$ at $(x_i^b,y_i^b)$ and the positive constant $s_+$ measures the degree of orientational ordering at the boundary \cite{canevari2017order,yin2020construction}. In 3D regions, the boundary loss function associated with tangential boundary conditions is given by
\begin{equation}
     W\sum_{i=1}^{N_b}\left\Vert \Q(x_i^b,y_i^b)\vec{\nu}_i^b + \frac{s_+}{3}\vec{\nu}_i^b \right \Vert_2^2.
     \label{boundary loss}
\end{equation}
The boundary loss \eqref{boundary loss} vanishes if and only if the leading eigenvector of $\Q$ lies within and is free to rotate in the tangential plane of $\partial \Omega$. The 3D homeotropic boundary conditions are $\Q_{bc}(x_i^b,y_i^b)=s_+(\vec{v}_i^b\times \vec{v}_i^b-\frac{\I_3}{3})$. 

% \begin{figure}
%         \centering
%    \captionsetup{justification=raggedright}     \includegraphics[width=0.7\linewidth]{Q_net.pdf}
%     \caption{The neural network approach of $\Q(x,y)$. In the 3D case, the input is $(x,y,z)$.}
%     \label{figure: Qnet}
% \end{figure}
\begin{figure}
    \centering
    \captionsetup{justification=raggedright}    \includegraphics[width=0.99\linewidth]{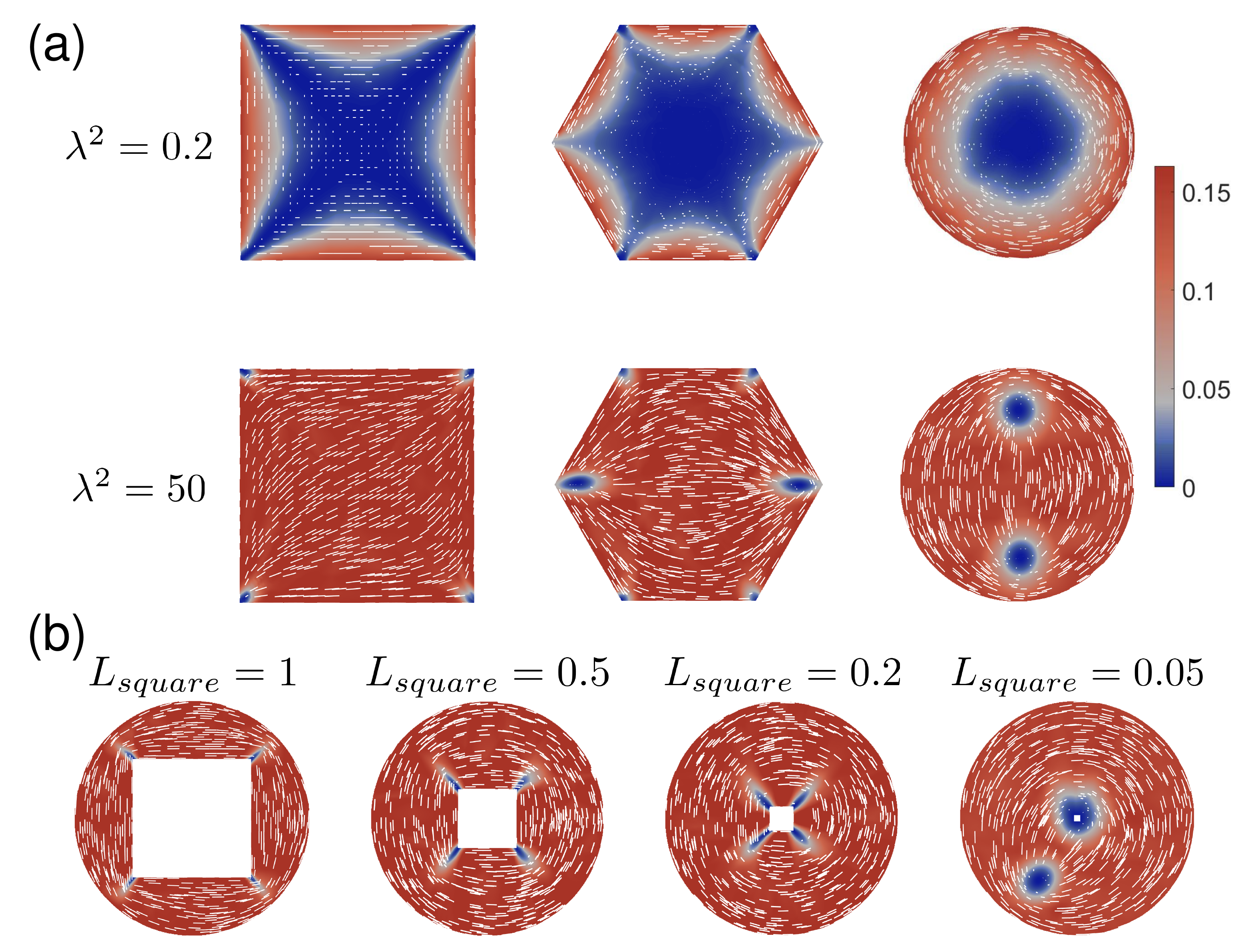}
    \caption{(a) NN-Tensor model applied to polygons for  $\kappa=7.243$. The defects are the isotropic blue regions. (b) NN-Tensor model applied to a square-holed disc for $\lambda^2=50,\kappa=7.243$. $L_{square}$ is the edge length of the inner square. The color bar labels $q_1^2+q_2^2=s^2/4$, and the white lines correspond to the nematic director, defined by the leading eigenvector of $\Q$.}
    \label{figure: confinement_2D_prl}
\end{figure}
In \cite{tsakonas2007multistable,han2020reduced}, the authors study NLC systems within regular $K$ polygons, in the 2D LdG framework, with tangent boundary conditions. In \cite{tsakonas2007multistable}, a 2D planar bistable nematic device is experimentally manufactured using quasi-2D square wells. For small $\lambda$, the LdG theory predicts a unique energy minimizer with an isolated defect (isotropic point) at the polygon centre, referred to as the Ring Solution \cite{han2020reduced}. For large $\lambda$, the LdG systems have $[K/2]$ classes of distinct energy minimizers with a pair of defects pinned at the vertices for a $K$-regular polygon \cite{han2020reduced}. The defects migrate to the interior as $K\rightarrow \infty$ and the $K$-polygon approaches a  disc \cite{hu2016disclination}. 
%The cases of an equilateral triangle and square domain need to be treated separately. 
In Figure \ref{figure: confinement_2D_prl}(a), we use the 2D NN-Tensor model to compute energy minimizers of \eqref{eq: Energy 2D inhomogeneous} on a square domain, hexagon and disc, for two representative values of $\lambda^2$, and the results are in agreement with the predictions in \cite{tsakonas2007multistable,han2020reduced, hu2016disclination}. 

In  Figure \ref{figure: confinement_2D_prl}(b), we apply the NN-Tensor model to a non-regular domain: square-holed unit discs, where the rescaled edge length of the inner square is $L_{square}$.  For a large hole with $L_{square}=1$, four point defects are pinned at the corners. As $L_{square}$ decreases, these four point defects merge and eventually separate into two +1/2 point defects. Notably, one point defect is pinned near the hole, indicating that holes can control defect locations. Similar phenomena are also found in self-assembly mechanisms in nematic emulsions \cite{zhou2019degenerate, senyuk2021transformation}. 

\begin{figure}
    \centering
    \captionsetup{justification=raggedright}   \includegraphics[width=0.95\linewidth]{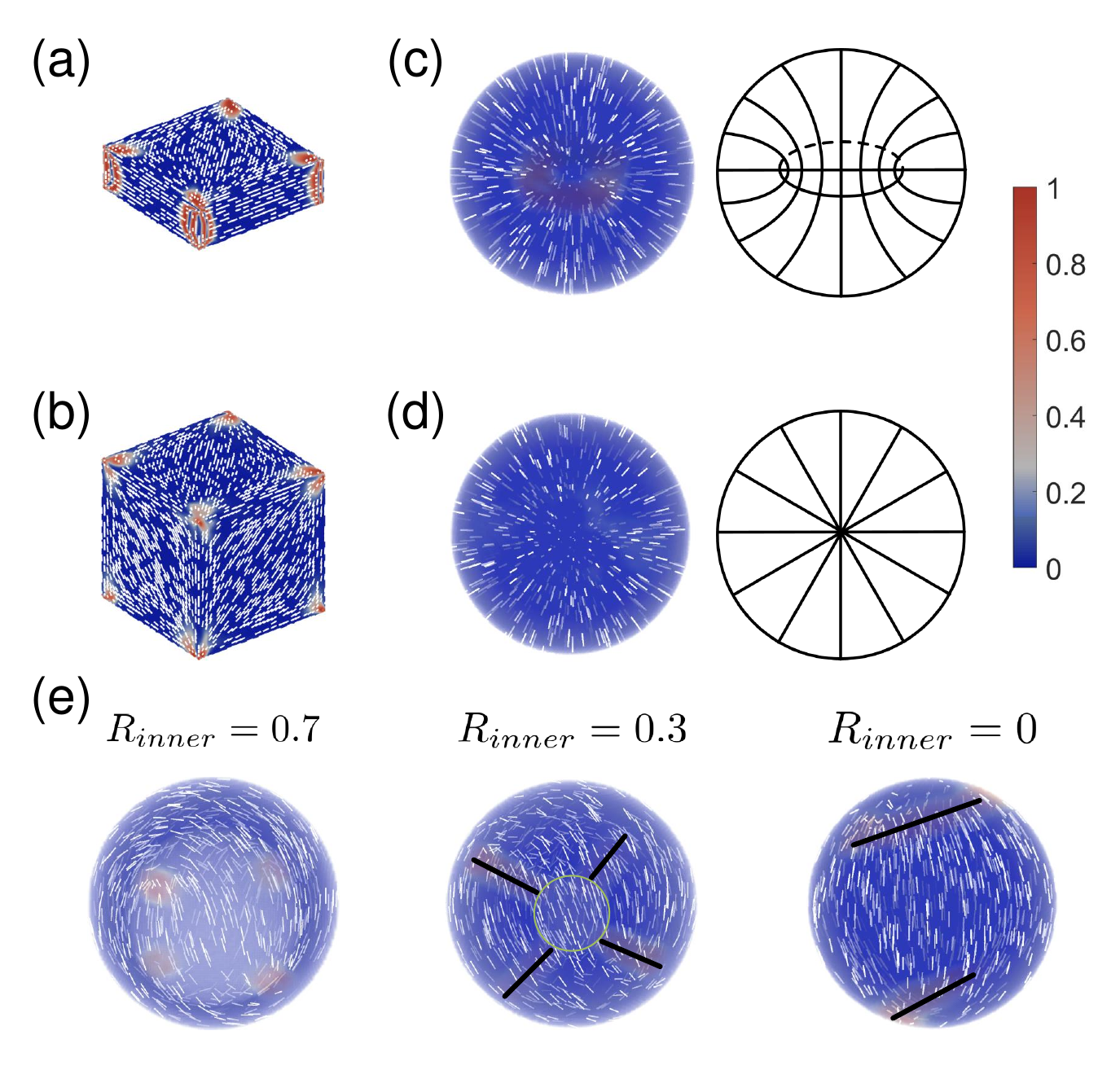}
    \caption{(a-b) NN-Tensor model applied to cuboids for  $\lambda^2=50, \kappa=12$ with $h=0.3$ and $h=1$, respectively. (c-d) NN-Tensor model applied to 3D balls with homeotropic bondary conditions, for $\lambda^2=30, \kappa=12$ and $\lambda^2=1, \kappa=12$, respectively. (e) NN-Tensor model applied to two asymmetric shells for $\lambda^2=30, \kappa=12$ with $R_{inner}=0.7,0.3,0$, respectively. The color bar labels the biaxial parameter  $\beta(\Q)=1-6\frac{(\text{Tr}(\Q^3))^2}{(\text{Tr}(\Q^2))^3}$.}
    \label{figure: 3D confined configurations}
\end{figure}

In \cite{shi2024multistability}, we study NLC configurations within 3D cuboids, with tangent boundary conditions, in the 3D LdG framework. For shallow cuboids, the minimizer is an almost $z$-invariant Diagonal (D) state which is consistent with the experimental observation in \cite{tsakonas2007multistable} i.e. the nematic director aligns along one of the square diagonals on the bottom cross-section and this profile is $z$-invariant along the height of the cuboid. For the cube, the LdG minimizer has a D-type profiles on all six faces, with defects at the corners. In Figure \ref{figure: 3D confined configurations} (a-b), we apply the 3D NN-Tensor model to the same problem and recover the LdG and experimental predictions. %we also have these two stable states with NN-Tensor model. 

We next consider a 3D ball with homeotropic boundary conditions, a very well-studied problem in the LdG framework. We apply the 3D NN-Tensor model to this problem and recover the celebrated radial hedgehog soluion with a central point defect, and the biaxial torus solution with a ring disclination, for various domain sizes as in \cite{mkaddem2000fine} and \cite{hu2016disclination}. Finally, in Figure \ref{figure: 3D confined configurations}(e), we consider a more complex 3D region between two asymmetric shells, with tangential boundary conditions in \eqref{boundary loss}, akin to the experimental settings in \cite{liang2011nematic,noh2020PRR,lopez2011nematic}. We fix the radius of the outer spherical shell, $R_{outer}=1$, with its center at $(0,0,0)$, and place the center of the inner spherical shell at $(0,0,-0.1)$. We then vary the radius of the inner spherical shell, $R_{inner}$. With a relatively thin shell, four $+1/2$ defects are experimentally observed \cite{liang2011nematic,lopez2011nematic}, which is predicted by our 3D NN-Tensor model at $R_{inner}=0.7$. For $R_{inner}=0$, the minimizer of \eqref{eq: Energy 2D inhomogeneous} has four point defects on the surface. These point defects are at the end-points of two non-intersecting  $1/2$ disclination lines. For $R_{inner}=0.3$,  the minimizer of \eqref{eq: Energy 2D inhomogeneous} exhibits four disclination lines connecting the inner and outer surfaces, demonstrating the crucial role of inner voids in controlling the locations and multiplicity of defects. 

%Conclusion
In conclusion, we have formulated a computationally efficient NN-Tensor model for NLCs, utilizing a neural network to map the macroscopic $\Q$-tensor to the microscopic M-S energy. As a result, the NN-Tensor model achieves energy precision comparable to the molecular model, accurately captures the I-N phase transition, and incorporates physical parameters derived directly from molecular information--capabilities beyond the conventional LdG tensor model.
We further embed the NN-Tensor model into the loss function of the second neural network, and stable NLC configurations are obtained by training this neural network in a domain-free and mesh-free manner, making it particularly suitable for complex domains of experimental interest.
%The NN-Tensor model not only successfully predicts a number of experimentally observed and theoretically expected NLC configurations in regular confinements, but also conveniently predicts new configurations in complex confinements. 

We use the NN-Tensor model to represent the bulk energy alone, because the elastic energy is derived analytically in \cite{han2015microscopic}. It should be noted that the NN-Tensor model can be extended to represent the elastic energy too. Furthermore, the NN-Tensor scheme is highly flexible and can be applied to other LC phases, such as the smectic phase \cite{han2015microscopic,mei2015molecular}, to other molecular shapes like bent-core molecules \cite{xu2018tensor} and to other physical systems described by the Landau theory of phase transitions, such as magnetic materials and superconductors \cite{hohenberg2015introduction}. 
%The proposed approach could be generalized to the other physical systems described by the Landau theory of phase transitions, such as magnetic materials and superconductors \cite{hohenberg2015introduction}.

%The NN-Tensor model is domain-free and mesh-free and has a $10^-3$ numerical error, which is acceptable for qualitative predictions. The NN-Tensor model predictions qualitatively agree with the conventional continuum LdG predictions. 

We thank Prof. Patrick E. Farrell for helpful discussions. This work is supported by the National Natural Science Foundation of China (No. 12225102, T2321001, and 12288101) and the Royal Society Newton Advanced Fellowship NAF/R1/180178 awarded to AM and LZ. 

%The methodology and all numerical simulations have been carried out by BS and LZ. AM contributed to the theoretical aspects, the model, and the benchmark examples when BS visited AM in August 2023 and January 2024.

\bibliography{references}

\end{document}